\documentclass[preprint]{elsarticle}
\usepackage{lineno,hyperref}
\modulolinenumbers[5]
\journal{}
\usepackage{setspace}

\usepackage{geometry}

\newgeometry{vmargin={25mm}, hmargin={20mm,20mm}}
\usepackage{color}
\usepackage{amsmath}

\begin{document}

\begin{frontmatter}

\title{A SIR model assumption for the spread of COVID-19 in different communities}

\author[IC_mainaddress]{Ian Cooper}

\author[AM_CA_mainaddress]{Argha Mondal\corref{AMcorrespondingauthor}}
\cortext[AMcorrespondingauthor]{Corresponding author}
\ead{arghamondalb1@gmail.com}

\author[AM_CA_mainaddress]{Chris G. Antonopoulos}

\address[IC_mainaddress]{School of Physics, The University of Sydney, Sydney, Australia}
\address[AM_CA_mainaddress]{Department of Mathematical Sciences, University of Essex, Wivenhoe Park, UK}

\begin{abstract}
In this paper, we study the effectiveness of the modelling approach on the pandemic due to the spreading of the novel COVID-19 disease and develop a susceptible-infected-removed (SIR) model that provides a theoretical framework to investigate its spread within a community. Here, the model is based upon the well-known susceptible-infected-removed (SIR) model with the difference that a total population is not defined or kept constant per se and the number of susceptible individuals does not decline monotonically. To the contrary, as we show herein, it can be increased in surge periods! In particular, we investigate the time evolution of different populations and monitor diverse significant parameters for the spread of the disease in various communities, represented by countries and the state of Texas in the USA. The SIR model can provide us with insights and predictions of the spread of the virus in communities that the recorded data alone cannot. Our work shows the importance of modelling the spread of COVID-19 by the SIR model that we propose here, as it can help to assess the impact of the disease by offering valuable predictions. Our analysis takes into account data from January to June, 2020, the period that contains the data before and during the implementation of strict and control measures. We propose predictions on various parameters related to the spread of COVID-19 and on the number of susceptible, infected and removed populations until September 2020. By comparing the recorded data with the data from our modelling approaches, we deduce that the spread of COVID-19 can be under control in all communities considered, if proper restrictions and strong policies are implemented to control the infection rates early from the spread of the disease.
\end{abstract}


\end{frontmatter}


\section{Introduction}
In December 2019, a novel strand of Coronavirus (SARS-CoV-2) was identified in Wuhan, Hubei Province, China causing a severe and potentially fatal respiratory syndrome, i.e., COVID-19. Since then, it has become a pandemic declared by World Health Organization (WHO) on March 11, which has spread around the globe \cite{Who2019,Wu2020,Novel19,Tang2020,Kraemer2020}. WHO published in its website preliminary guidelines with public health care for the countries to deal with the pandemic \cite{WHO19}. Since then, the infectious disease has become a public health threat. Italy and USA are severely affected by COVID-19 \cite{Song2020,Wang2020,Dan2020}. Millions of people are forced by national governments to stay in self-isolation and in difficult conditions. The disease is growing fast in many countries around the world. In the absence of availability of a proper medicine or vaccine, currently social distancing, self-quarantine and wearing a face mask have been emerged as the most widely-used strategy for the mitigation and control of the pandemic.

In this context, mathematical models are required to estimate disease transmission, recovery, deaths and other significant parameters separately for various countries, that is for different, specific regions of high to low reported cases of COVID-19. Different countries have already taken precise and differentiated measures that are important to control the spread of the disease. However, still now, important factors such as population density, insufficient evidence for different symptoms, transmission mechanism and unavailability of a proper vaccine, makes it difficult to deal with such a highly infectious and deadly disease, especially in high population density countries such as India \cite{Ranjan2020,Pulla2020,MOH2020}. Recently, many research articles have adopted the modelling approach, using real incidence datasets from affected countries and, have investigated different characteristics as a function of various parameters of the outbreak and the effects of intervention strategies in different countries, respective to their current situations.

It is imperative that mathematical models are developed to provide insights and make predictions about the pandemic, to plan effective control strategies and policies \cite{Scarpino2019,Chinazzi2020,Rud2020}. Modelling approaches \cite{Wang2020,Kuchar2020,Yang2020,Fan2020,Xue2020,Post2020,Li2020} are helpful to understand and predict the possibility and severity of the disease outbreak and, provide key information to determine the intensity of COVID-19 disease intervention. The susceptible-infected-removed (SIR) model and its extended modifications \cite{Heth1989,Heth2000,Heth2009,Weiss2013}, such as the extended-susceptible-infected-removed (eSIR) mathematical model in various forms have been used in previous studies \cite{Amaro2020,Cala2020,Ndairou2020} to model the spread of COVID-19 within communities.

Here, we propose the use of a novel SIR model with different characteristics. One of the major assumptions of the classic SIR model is that there is a homogeneous mixing of the infected and susceptible populations and that the total population is constant in time. In the classic SIR model, the susceptible population decreases monotonically towards zero. However, these assumptions are not valid in the case of the spread of the COVID-19 virus, since new epicentres spring up around the globe at different times. To account for this, the SIR model that we propose here does not consider the total population and takes the susceptible population as a variable that can be adjusted at various times to account for new infected individuals spreading throughout a community, resulting in an increase in the susceptible population, i.e., to the so-called surges. The SIR model we introduce here is given by the same simple system of three ordinary differential equations (ODEs) with the classic SIR model and can be used to gain a better understanding of how the virus spreads within a community of variable populations in time, when surges occur. Importantly, it can be used to make predictions of the number of infections and deaths that may occur in the future and provide an estimate of the time scale for the duration of the virus within a community. It also provides us with insights on how we might lessen the impact of the virus, what is nearly impossible to discern from the recorded data alone! Consequently, our SIR model can provide a theoretical framework and predictions that can be used by government authorities to control the spread of COVID-19.

In our study, we used COVID-19 datasets from \cite{Corona} in the form of time-series, spanning January to June, 2020. In particular, the time series are composed of three columns which represent the total cases $I^{d}_{tot}$, active cases $I^d$ and Deaths $D^d$ in time (rows). These datasets were used to update parameters of the SIR model to understand the effects and estimate the trend of the disease in various communities, represented by China, South Korea, India, Australia, USA, Italy and the state of Texas in the USA. This allowed us to estimate the development of COVID-19 spread in these communities by obtaining estimates for the number of deaths $D$, susceptible $S$, infected $I$ and removed $R_m$ populations in time. Consequently, we have been able to estimate its characteristics for these communities and assess the effectiveness of modelling the disease.

The paper is organised as following: In Sec. \ref{sec_SIR_model}, we introduce the SIR model and discuss its various aspects. In Sec. \ref{sec_methodology_and_results}, we explain the approach we used to study the data in \cite{Corona} and in Sec. \ref{sec_results}, we present the results of our analysis for China, South Korea, India, Australia, USA, Italy and the state of Texas in the USA. Section \ref{sec_flattening_the_curve} discusses the implications of our study to the ``flattening the curve'' approach. Finally, in Sec. \ref{sec_conclusions}, we conclude our work and discuss the outcomes of our analysis and its connection to the evidence that has been already collected on the spread of COVID-19 worldwide.

\section{The SIR model that can accommodate surges in the susceptible population}\label{sec_SIR_model}
The world around us is highly complicated. For example, how a virus spreads, including the novel strand of Coronavirus (SARS-CoV-2) that was identified in Wuhan, Hubei Province, China, depends upon many factors, among which some of them are considered by the classic SIR model, which is rather simplistic and cannot take into consideration surges in the number of susceptible individuals. Here, we propose the use of a modified SIR model with characteristics, based upon the classic SIR model. In particular, one of the major assumptions of the classic SIR model is that there is a homogeneous mixing of the infected $I$ and susceptible $S$ populations and that the total population $N$ is constant in time. Also, in the SIR model, the susceptible population $S$ decreases monotonically towards zero. These assumptions however are not valid in the case of the spread of the COVID-19 virus, since new epicentres spring up around the globe at different times. To account for this, we introduce here a SIR model that does not consider the total population $N$, but rather, takes the susceptible population $S$ as a variable that can be adjusted at various times to account for new infected individuals spreading throughout a community, resulting in its increase. Thus, our model is able to accommodate surges in the number of susceptible individuals in time, whenever these occur and as evidenced by published data, such as those in \cite{Corona} that we consider here.

Our SIR model is given by the same, simple system of three ordinary differential equations (ODEs) with the classic SIR model that can be easily implemented and used to gain a better understanding of how the COVID-19 virus spreads within communities of variable populations in time, including the possibility of surges in the susceptible populations. Thus, the SIR model here is designed to remove many of the complexities associated with the real-time evolution of the spread of the virus, in a way that is useful both quantitatively and qualitatively. It is a dynamical system that is given by three coupled ODEs that describe the time evolution of the following three populations:
\begin{enumerate}
\item {{\it Susceptible individuals}, $S(t)$: These are those individuals who are not infected, however, could become infected. A susceptible individual may become infected or remain susceptible. As the virus spreads from its source or new sources occur, more individuals will become infected, thus the susceptible population will increase for a period of time (surge period).}
\item{ {\it Infected individuals}, $I(t)$: These are those individuals who have already been infected by the virus and can transmit it to those individuals who are susceptible. An infected individual may remain infected, and can be removed from the infected population to recover or die.}
\item{{\it Removed individuals}, $R_m(t)$: These are those individuals who have recovered from the virus and are assumed to be immune, $R_m(t)$ or have died, $D(t)$.}
\end{enumerate}
Furthermore, it is assumed that the time scale of the SIR model is short enough so that births and deaths (other than deaths caused by the virus) can be neglected and that the number of deaths from the virus is small compared with the living population.

Based on these assumptions and concepts, the rates of change of the three populations are governed by the following system of ODEs, what constitutes our SIR model
\begin{equation}
\begin{aligned}
\frac{dS(t)}{dt}&=-aS(t)I(t),\\
\frac{dI(t)}{dt}&=aS(t)I(t) - bI(t),\\
\frac{d{R_m}(t)}{dt}&=bI(t),
\label{SIR_model_ODEs}
\end{aligned}
\end{equation}
where $a$ and $b$ are real, positive, parameters of the initial exponential growth and final exponential decay of the infected population $I$.

It has been observed that in many communities, a spike in the number of infected individuals, $I$, may occur, which results in a surge in the susceptible population, $S$, recorded in the COVID-19 datasets \cite{Corona}, what amounts to a secondary wave of infections. To account for such a possibility, $S$ in the SIR model \eqref{SIR_model_ODEs}, can be reset to $S_{surge}$ at any time $t_s$ that a surge occurs, and thus it can accommodate multiple such surges if recorded in the published data in \cite{Corona}, what distinguishes it from the classic SIR model.

The evolution of the infected population $I$ is governed by the second ODE in system \ref{SIR_model_ODEs}, where $a$ is the transmission rate constant and $b$ the removal rate constant. We can define the basic effective reproductive rate $R_e=aS(t)/b$, as the fate of the evolution of the disease depends upon it. If $R_e$ is smaller than one, the infected population $I$ will decrease monotonically to zero and if greater than one, it will increase, i.e., if $\frac{dI(t)}{dt}<0\Rightarrow R_e<1$ and if $\frac{dI(t)}{dt} > 0\Rightarrow R_e>1$. Thus, the effective reproductive rate $R_e$ acts as a threshold that determines whether an infectious disease will die out quickly or will lead to an epidemic.

At the start of an epidemic, when $R_e>1$ and $S \approx1$, the rate of infected population is described by the approximation $\frac{{dI(t)}}{{dt}}\approx\left({a-b} \right)I(t)$ and thus, the infected population $I$ will initially increase exponentially according to $I(t)=I(0)\,{e^{(a-b)t}}$. The infected population will reach a peak when the rate of change of the infected population is zero, $dI(t)/dt=0$, and this occurs when  $R_e=1$. After the peak, the infected population will start to decrease exponentially, following $I(t) \propto{e^{-bt}}$. Thus, eventually (for $t\rightarrow\infty$), the system will approach $S\to0$ and $I\to0$. Interestingly, the existence of a threshold for infection is not obvious from the recorded data, however can be discerned from the model. This is crucial in identifying a possible second wave where a sudden increase in the susceptible population $S$ will result in $R_e>1$, and to another exponential growth of the number of infections $I$.

\section{Methodology}\label{sec_methodology_and_results}

The data in \cite{Corona} for China, South Korea, India, Australia, USA, Italy and the state of Texas (communities) are organised in the form of time-series where the rows are recordings in time (from January to June, 2020), and the three columns are, the total cases $I^d_{tot}$ (first column), number of infected individuals $I^d$ (second column) and deaths $D^d$ (third column). Consequently, the number of removals can be estimated from the data by
\begin{equation}\label{removals_data_equation}
R^d_m=I^d_{tot}-I^d-D^d.
\end{equation}
Since we want to adjust the numerical solutions to our proposed SIR model \eqref{SIR_model_ODEs} to the recorded data from \cite{Corona}, for each dataset (community), we consider initial conditions in the interval $[0,1]$ and scale them by a scaling factor $f$ to fit the recorded data by visual inspection. In particular, the initial conditions for the three populations are set such that $S(0)=1$ (i.e., all individuals are considered susceptible initially), $I(0)=R_m(0)=I^d_{max}/f<1$, where $I^d_{max}$ is the maximum number of infected individuals $I^d$. Consequently, the parameters $a$, $b$, $f$ and $I^d_{max}$ are adjusted manually to fit the recorded data as best as possible, based on a trial-and-error approach and visual inspections. A preliminary analysis using non-linear fittings to fit the model to the published data \cite{Corona} provided at best inferior results to those obtained in this paper using our trial-and-error approach with visual inspections, in the sense that the model solutions did not follow as close the published data, what justifies our approach in the paper. A prime reason for this is that the published data (including those in \cite{Corona} we are using here) are data from different countries that follow different methodologies to record them, with not all infected individuals or deaths accounted for.

In this context, $S$, $I$ and $R_m\geq0$ at any $t\geq0$. System \eqref{SIR_model_ODEs} can be solved numerically to find how the scaled (by $f$) susceptible $S$, infected $I$ and removed $R_m$ populations (what we call model solutions) evolve with time, in good agreement with the recorded data. In particular, since this system is simple with well-behaved solutions, we used the first-order Euler integration method to solve it numerically, and a time step $h=200/5000=0.04$ that corresponds to a final integration time $t_f$ of 200 days since January, 2020. This amounts to double the time interval in the recorded data in \cite{Corona} and allows for predictions for up to 100 days after January, 2020.

Obviously, what is important when studying the spread of a virus is the number of deaths $D$ and recoveries $R$ in time. As these numbers are not provided directly by the SIR model \eqref{SIR_model_ODEs}, we estimated them by first, plotting the data for deaths $D^d$ vs the removals $R^d_m$, where $R^d_m=D^d+R^d=I^d_{tot}-I^d$ and then  fitting the plotted data  with the nonlinear function
\begin{equation}\label{nonlinear_fitting_function_Dd_Rmd}
D={D_0}\,\left({1-{e^{-k{R_m}}}}\right),
\end{equation}
where $D_0$ and $k$ are constants estimated by the non-linear fitting. The function is expressed in terms of only model values and is fitted to the curve of the data. Thus, having obtained $D$ from the non-linear fitting, the number of recoveries $R$ can be described in time by the simple observation that it is given by the scaled removals, $R_m$ from the SIR model \eqref{SIR_model_ODEs}, less the number of deaths, $D$ from Eq. \eqref{nonlinear_fitting_function_Dd_Rmd},
\begin{equation}\label{recoveries_equation}
R=R_m-D.
\end{equation}

\section{Results}\label{sec_results}

The rate of increase in the number of infections depends on the product of the number of infected and susceptible individuals. An understanding of the system of Eqs. \eqref{SIR_model_ODEs} explains the staggering increase in the infection rate around the world. Infected people traveling around the world has led to the increase in infected numbers and this results in a further increase in the susceptible population \cite{Chinazzi2020}. This gives rise to a positive feedback loop leading to a very rapid rise in the number of active infected cases. Thus, during a surge period, the number of susceptible individuals increases and as a result, the number of infected individuals increases as well. For example, as of 1 March, 2020, there were 88 590 infected individuals and by 3 April, 2020, this number had grown to a staggering 1 015 877 \cite{Corona}. 

Understanding the implications of what the system of Eqs. \eqref{SIR_model_ODEs} tells us, the only conclusion to be drawn using scientific principles is that drastic action needs to be taken as early as possible, while the numbers are still low, before the exponential increase in infections starts kicking in. For example, if we consider the results of policies introduced in the UK to mitigate the spread of the disease, there were 267 240 total infections and 37 460 deaths by 27 May and in the USA, 1 755 803 and 102 107, total infections and deaths, respectively. Thus, even if one starts with low numbers of infected individuals, the number of infections will at first grow slowly and then, increase approximately exponentially, then taper off until a peak is reached. Comparing these results for the UK and USA with those for South Korea, where steps were taken immediately to reduce the susceptible population, there were 11 344 total infections and 269 deaths by 27 May. The number of infections in China reached a peak about 16 February, 2020. The government took extreme actions with closures, confinement, social distancing, and people wearing masks. This type of action produces a decline in the number of infections and susceptible individuals. If the number of susceptible individuals does not decrease, then the number of infections just gets increased rapidly. As at this moment, there is no effective vaccine developed, the only way to reduce the number of infections is to reduce the number of individuals that are susceptible to the disease. Consequently, the rate of infection tends to zero only if the susceptible population goes to zero. 

Here, we have applied the SIR model \eqref{SIR_model_ODEs} considering data from various countries and the state of Texas in the USA provided in \cite{Corona}. Assuming the published data are reliable, the SIR model \eqref{SIR_model_ODEs} can be applied to assess the spread of the COVID-19 disease and predict the number of infected, removed and recovered populations and deaths in the communities, accommodating at the same time possible surges in the number of susceptible individuals. Figures \ref{Fig1}--\ref{Fig9} show the time evolution of the cumulative total infections $I_{tot}$, current infected individuals, $I$, recovered individuals, $R$, dead individuals, $D$, and normalized susceptible populations, $S$ for China, South Korea, India, Australia, USA, Italy and Texas in the USA, respectively. The crosses show the published data \cite{Corona} and the smooth lines, solutions and predictions from the SIR model. The cumulative total infections plots also show a curve for the initial exponential increase in the number of infections, where the number of infections doubles every five days. The figures also show predictions, and a summary of the SIR model parameters in \eqref{SIR_model_ODEs} and published data in \cite{Corona} for easy comparisons.

We start by analysing the data from China and then move on to the study of the data from South Korea, India, Australia, USA, Italy and Texas.

\subsection{China}

\begin{figure}
\centering
\includegraphics[width=14cm,height=9.5cm]{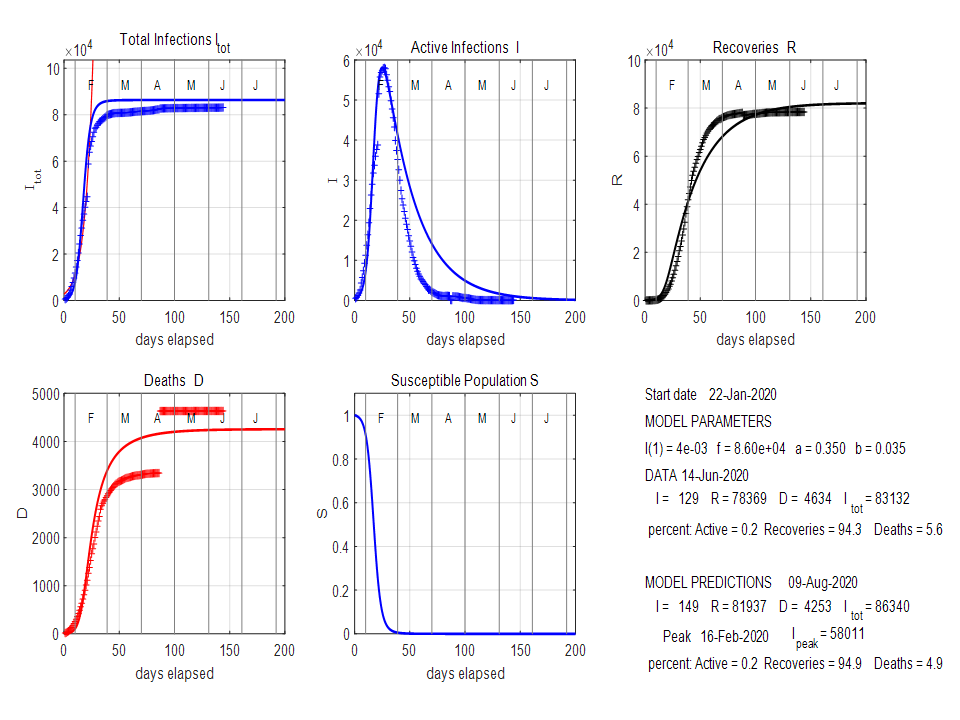}
\caption{China: Model predictions for the period from 22 January to 9 August, 2020 with data from January to June, 2020. The data show a discrete jump in deaths $D$ in mid-April.}\label{Fig1}
\end{figure}
\begin{figure}
\centering
\includegraphics[width=12cm,height=8.5cm]{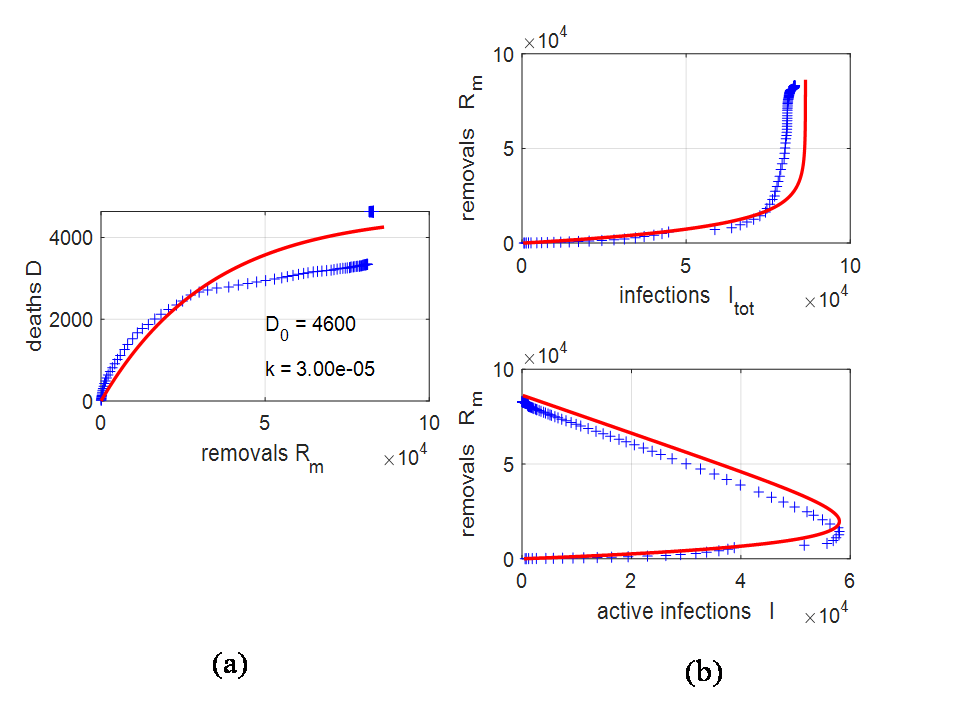}
\caption{China: (a) Nonlinear fitting with Eq. \eqref{nonlinear_fitting_function_Dd_Rmd} using a trial-and-error method to estimate the number of deaths, $D$ from the removed population, $R_m$ (see text for the details). (b) Plots of the number of removals, $R_m$ against the cumulative total infections $I_{tot}$ and current active cases $I$.}\label{Fig1A}
\end{figure}

The number of infections peaked in China about 16 February, 2020 and since then, it has slowly decreased. The decrease only occurs when the susceptible population numbers decrease and this decrease in susceptible numbers only occurred through the drastic actions taken by the Chinese government. China quarantined and confirmed potential patients, and restricted citizens' movements as well as international travel. Social distancing was widely practiced, and most of the people wore face masks. The actual numbers of infections have decreased at a greater rate than predicted by the SIR model (see Figs. \ref{Fig1} and \ref{Fig1A}). Our results in Figs. \ref{Fig1} and \ref{Fig1A} provide evidence that the Chinese government has done well in limiting the impact of the spread of COVID-19.

\subsection{South Korea}

\begin{figure}
	\centering
	\includegraphics[width=14cm,height=8.5cm]{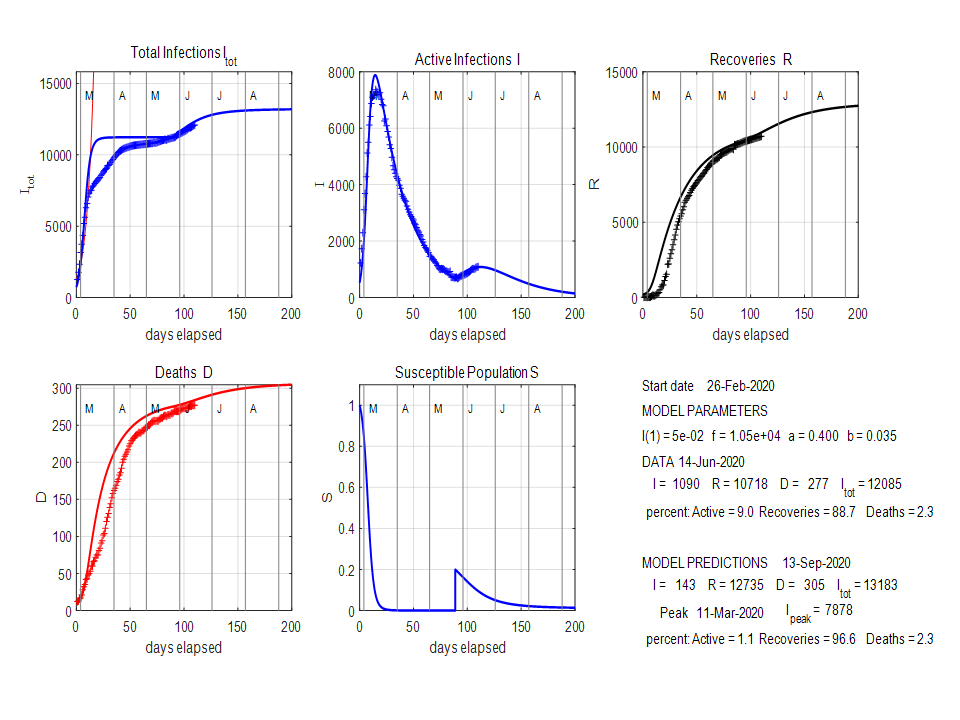}
	\caption{South Korea: Model predictions for the period from 26 February to 13 September, 2020 with data from February to June, 2020.}
	\label{Fig2}
\end{figure}
\begin{figure}
	\centering
	\includegraphics[width=12cm,height=8.5cm]{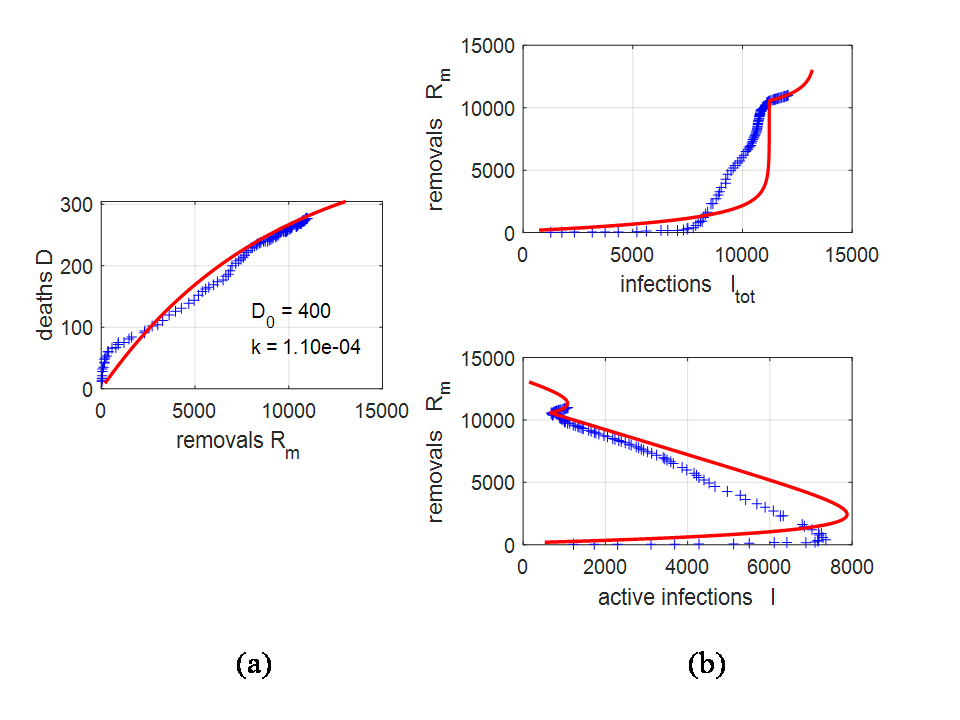}
	\caption{South Korea: (a) Nonlinear fitting with Eq. \eqref{nonlinear_fitting_function_Dd_Rmd} using a trial-and-error method to estimate the number of deaths, $D$ from the removed population, $R_m$ (see text for the details). (b) Plots of the number of removals, $R_m$ against the cumulative total infections $I_{tot}$ and current active cases $I$.} 
	\label{Fig2A}
\end{figure}

From the plots shown in Figs. \ref{Fig2} and \ref{Fig2A}, it is obvious that the South Korean government has done a wonderful job in controlling the spread of the virus. The country has implemented an extensive virus testing program. There has also been a heavy use of surveillance technology: closed-circuit television (CCTV) and tracking of bank cards and mobile phone usage, to identify who to test in the first place. South Korea has achieved a low fatality rate (currently one percent) without resorting to such authoritarian measures as in China. The most conspicuous part of the South Korean strategy is simple enough: implementation of repeated cycles of {\it test and contact trace} measures.

\subsection{India}

\begin{figure}
	\centering
	\includegraphics[width=14cm,height=8.5cm]{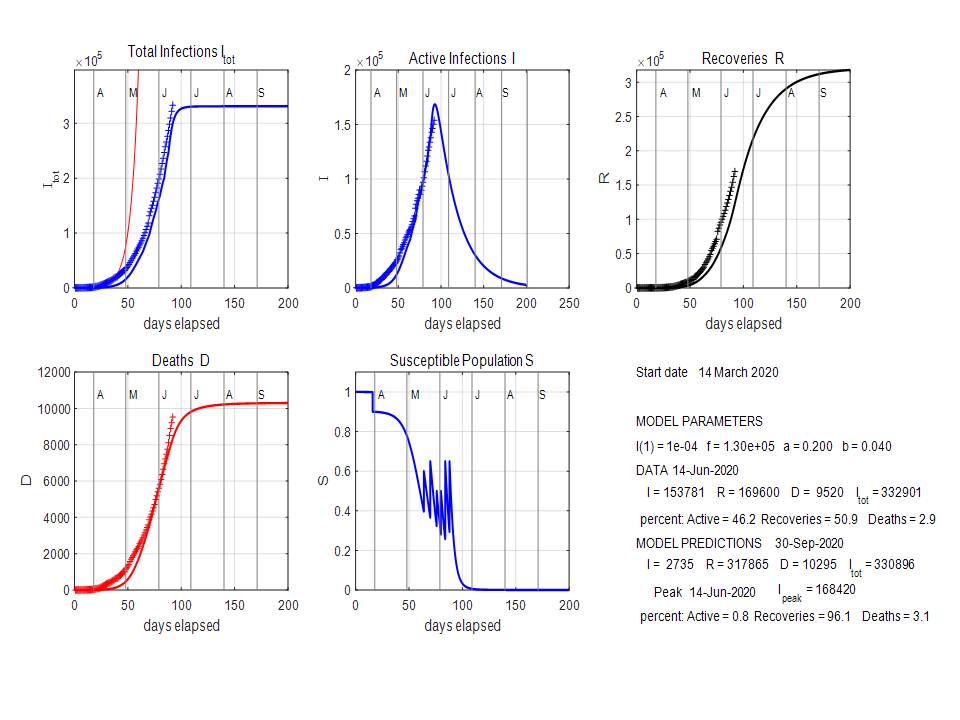}
	\caption{India: Model predictions for the period from 14 March to 30 September, 2020 with data from March to June, 2020.} 
	\label{Fig3}
\end{figure}
\begin{figure}
	\centering
	\includegraphics[width=12cm,height=8.5cm]{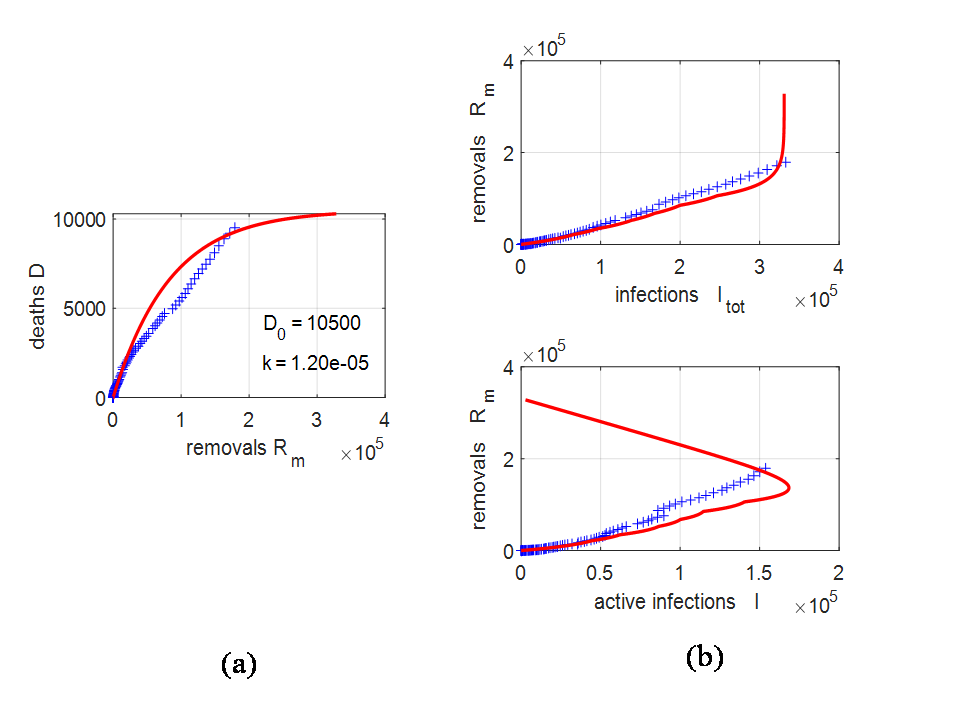}
	\caption{India: (a) Nonlinear fitting with Eq. \eqref{nonlinear_fitting_function_Dd_Rmd} using a trial-and-error method to estimate the number of deaths, $D$ from the removed population, $R_m$ (see text for the details). (b) Plots of the number of removals, $R_m$ against the cumulative total infections $I_{tot}$ and current active cases $I$.} 
	\label{Fig3A}
\end{figure}

To match the recorded data from India with predictions from the SIR model \eqref{SIR_model_ODEs}, it is necessary to include a number of surge periods, as shown in Fig. \ref{Fig3}. This is because the SIR model cannot predict accurately the peak number of infections, if the actual numbers in the infected population have not peaked in time. It is most likely the spread of the virus as of early June, 2020 is not contained and there will be an increasing number of total infections. However, by adding new surge periods, a higher and delayed peak can be predicted and compared with future data. In Fig. \ref{Fig3}, a consequence of the surge periods is that the peak is delayed and higher than if no surge periods were applied. The model predictions for the 30 September, 2020 including the surges are: 330 000 total infections,  700 active infections and 7 500 deaths, whereas if there were no surge periods, there would be 130 000 total infections, 700 active infections and 6 300 deaths, with the peak of 60 000, which is about 40\% of the current number of active cases occuring around 20 May 2020. Thus, the model can still give a rough estimate of future infections and deaths, as well as the time it may take for the number of infections to drop to safer levels, at which time restrictions can be eased, even without an accurate prediction in the peak in active infections (see Figs. \ref{Fig3} and \ref{Fig3A}).

\subsection{Australia}
\begin{figure}
	\centering
	\includegraphics[width=14cm,height=8.5cm]{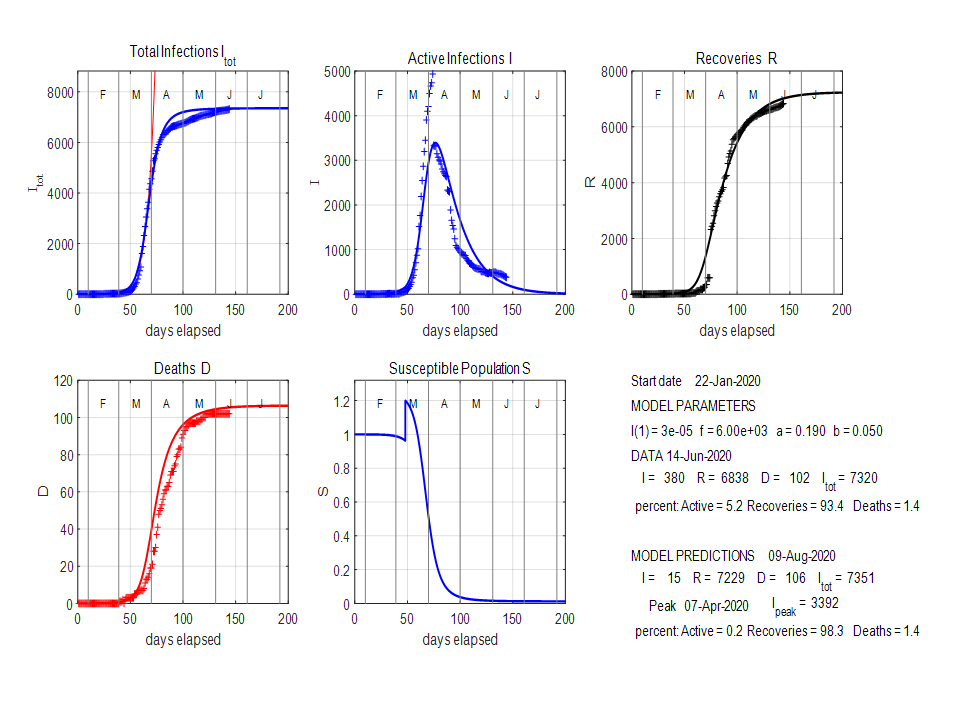}
	\caption{Australia: Model predictions for the period from 22 January to 9 August, 2020 with data from January to June, 2020.} 
	\label{Fig4}
\end{figure}
\begin{figure}
	\centering
	\includegraphics[width=12cm,height=8.5cm]{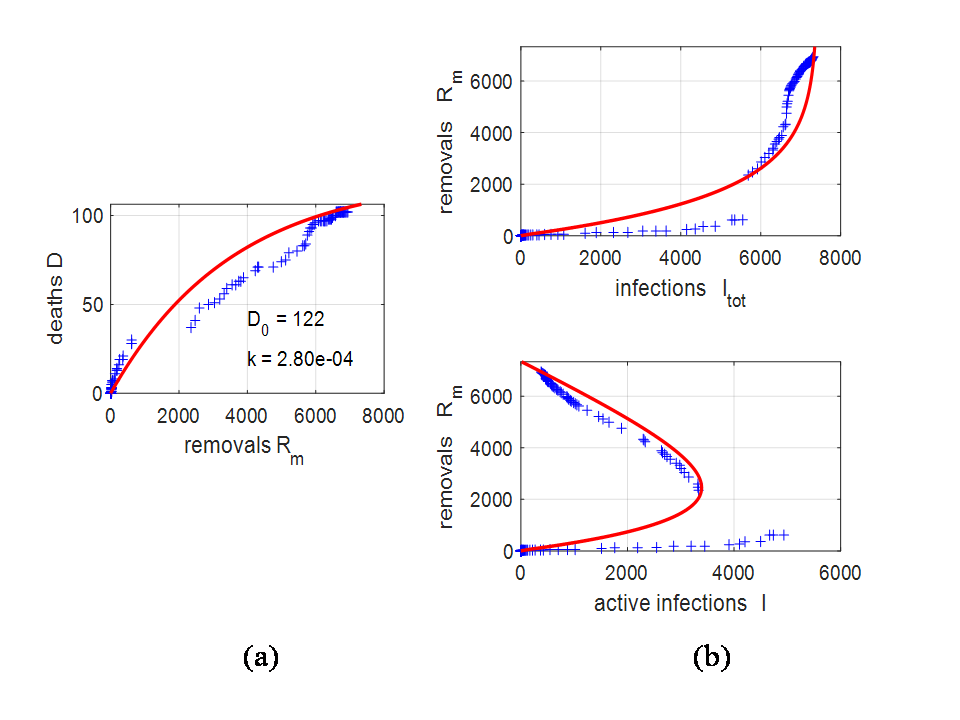}
	\caption{Australia: (a) Nonlinear fitting with Eq. \eqref{nonlinear_fitting_function_Dd_Rmd} using a trial-and-error method to estimate the number of deaths, $D$ from the removed population, $R_m$ (see text for the details). (b) Plots of the number of removals, $R_m$ against the cumulative total infections $I_{tot}$ and current active cases $I$.} 
	\label{Fig4A}
\end{figure}

A surge in the susceptible population was applied in early March, 2020 in the country. The surge was caused by 2 700 passengers disembarking from the Ruby Princes cruise ship in Sydney and then, returning to their homes around Australia. More than 750 passengers and crew have become infected and 26 died. Two government enquires have been established to investigate what went wrong. Also, at this time many infected overseas passengers arrived by air from Europe and the USA. The Australian government was too slow in quarantining arrivals from overseas.

From mid-March, 2020 until mid-May, 2020, the Australian governments introduced measures of testing, contact tracing, social distancing, staying at home policy, closure of many businesses and encouraging people to work from home. From Figs. \ref{Fig4} and \ref{Fig4A}, it can be observed that actions taken were successful as the actual number of infections declined in accord with the model predictions. There have been no further surge periods. From end of May, 2020, these restrictions are being removed in stages. The SIR model can be used when future data becomes available to see if the number of susceptible individuals starts to increase. If so, the model can accommodate this by introducing surge factors.

\subsection{USA}
\begin{figure}
	\centering
	\includegraphics[width=14cm,height=8.5cm]{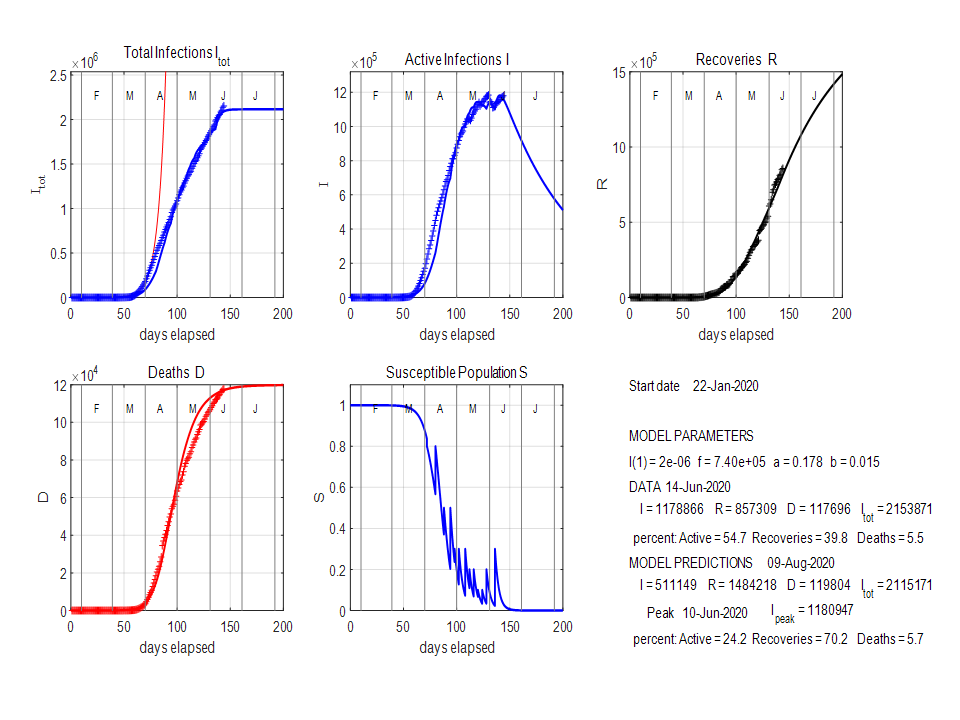}
	\caption{USA: Model predictions for the period from 22 January to 9 August, 2020 with data from January to June, 2020.} 
	\label{Fig5}
\end{figure}
\begin{figure}
	\centering
	\includegraphics[width=12cm,height=8.5cm]{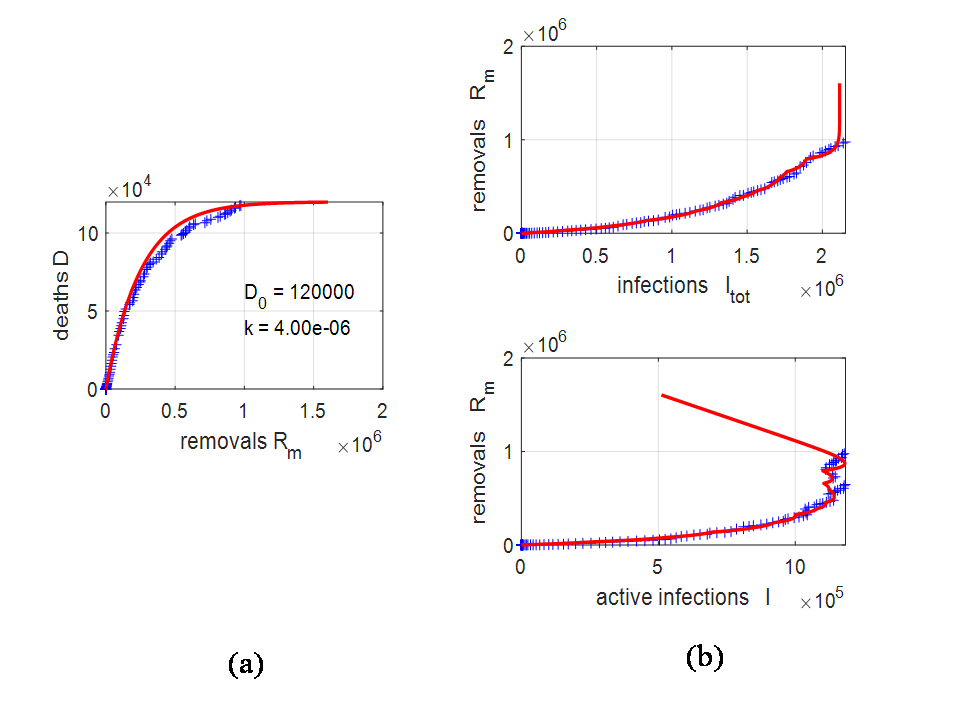}
	\caption{USA: (a) Nonlinear fitting with Eq. \eqref{nonlinear_fitting_function_Dd_Rmd} using a trial-and-error method to estimate the number of deaths, $D$ from the removed population, $R_m$ (see text for the details). (b) Plots of the number of removals, $R_m$ against the cumulative total infections $I_{tot}$ and current active cases $I$.} 
	\label{Fig5A}
\end{figure}

As of early June, 2020, the peak number of infections has not been reached. When a peak in the data is not reached, it is more difficult to fit the model predictions to the data. In the model, it is necessary to add a few surge periods. This is because new epicentres of the virus arose at different times. The virus started spreading in Washington State, followed by California, New York, Chicago and the southern states of the USA. The need to add surge periods shows clearly that the spread of the virus is not under control.

In the USA, by the end of May, 2020, the number of active infected cases has not yet peaked and the cumulative total number of infections keeps getting bigger. This can be accounted for in the SIR model by considering how the susceptible population changes with time in May. During that time, to match the data to the model predictions, surge periods were used where the normalized susceptible population $S$ was reset to $0.2$ every four days. What is currently happening in the USA is that as susceptible individuals become infected, their population decreases, with these infected individuals mixing with the general population, leading to an increase in the susceptible population. This is shown in the model by the variable for the susceptible population, $S$, varying from about $0.06$ to $0.20$, repeatedly during May. Until this vicious cycle is broken, the cumulative total infected population will keep growing at a steady rate and not reach an almost steady-state. The fluctuating normalized susceptible variable provides clear evidence that government authorities do not have the spread of the virus under control (see Figs. \ref{Fig5} and \ref{Fig5A}).

\subsection{Texas}

\begin{figure}
	\centering
	\includegraphics[width=14cm,height=8.5cm]{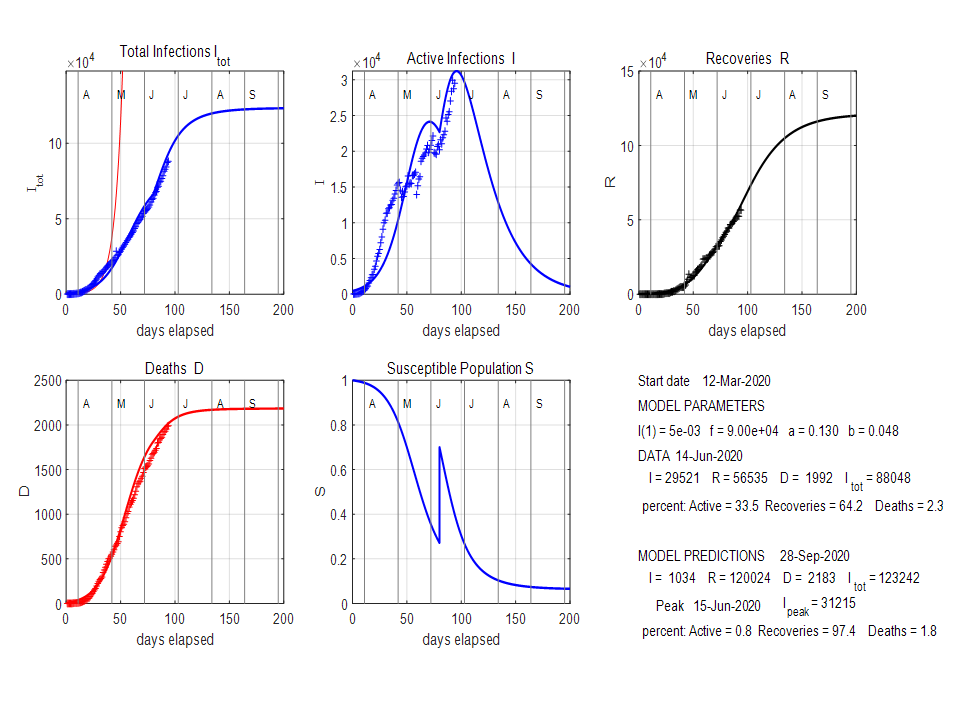}
	\caption{Texas: Model predictions for the period from 12 March to 28 September, 2020 with data from March to June, 2020.} 
	\label{Fig6}
\end{figure}
\begin{figure}
	\centering
	\includegraphics[width=12cm,height=8.5cm]{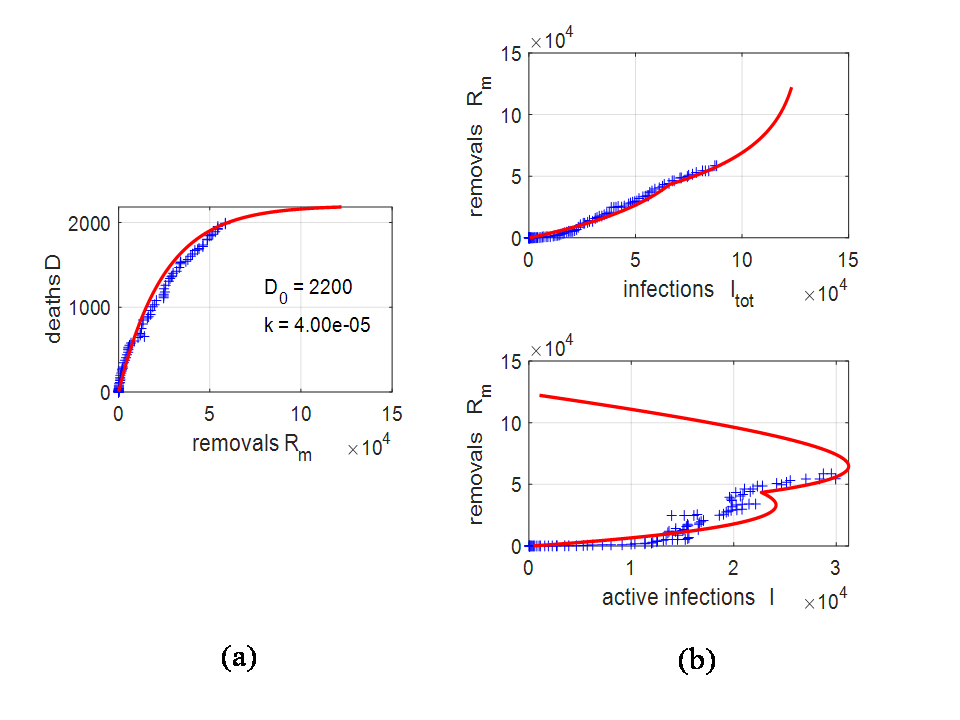}
	\caption{Texas: (a) Nonlinear fitting with Eq. \eqref{nonlinear_fitting_function_Dd_Rmd} using a trial-and-error method to estimate the number of deaths, $D$ from the removed population, $R_m$ (see text for the details). (b) Plots of the number of removals, $R_m$ against the cumulative total infections $I_{tot}$ and current active cases $I$.} 
	\label{Fig6A}
\end{figure}

\begin{figure}
	\centering
	\includegraphics[width=14cm,height=8.5cm]{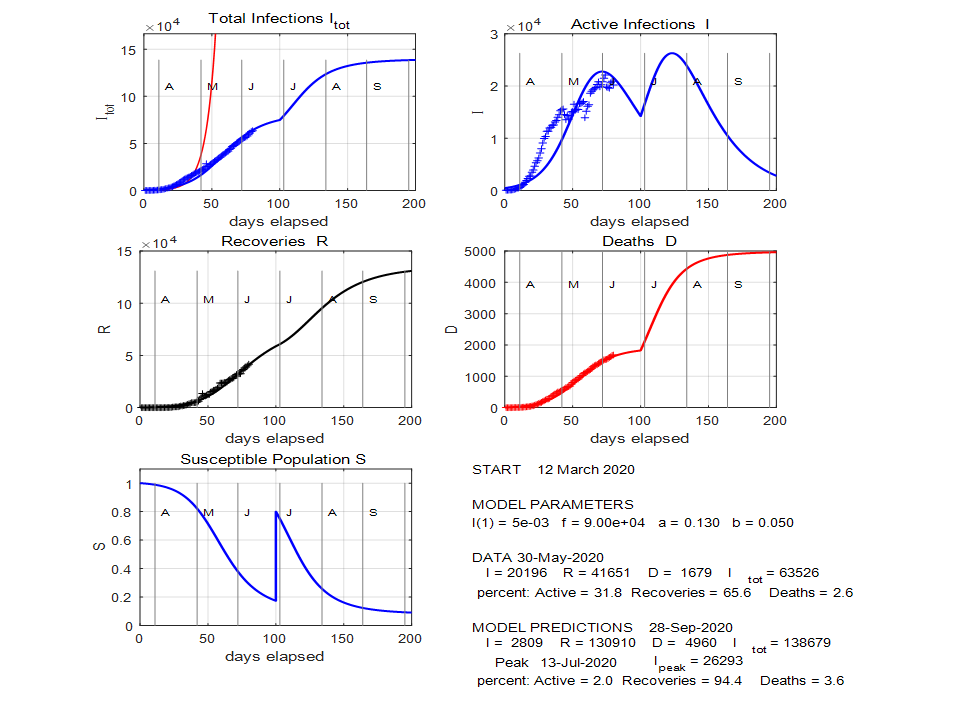}
	\caption{Texas: Model predictions with a surge period occurring at the end of June, 2020.} 
	\label{Fig7}
\end{figure}

\begin{figure}
	\centering
	\includegraphics[width=12cm,height=8.5cm]{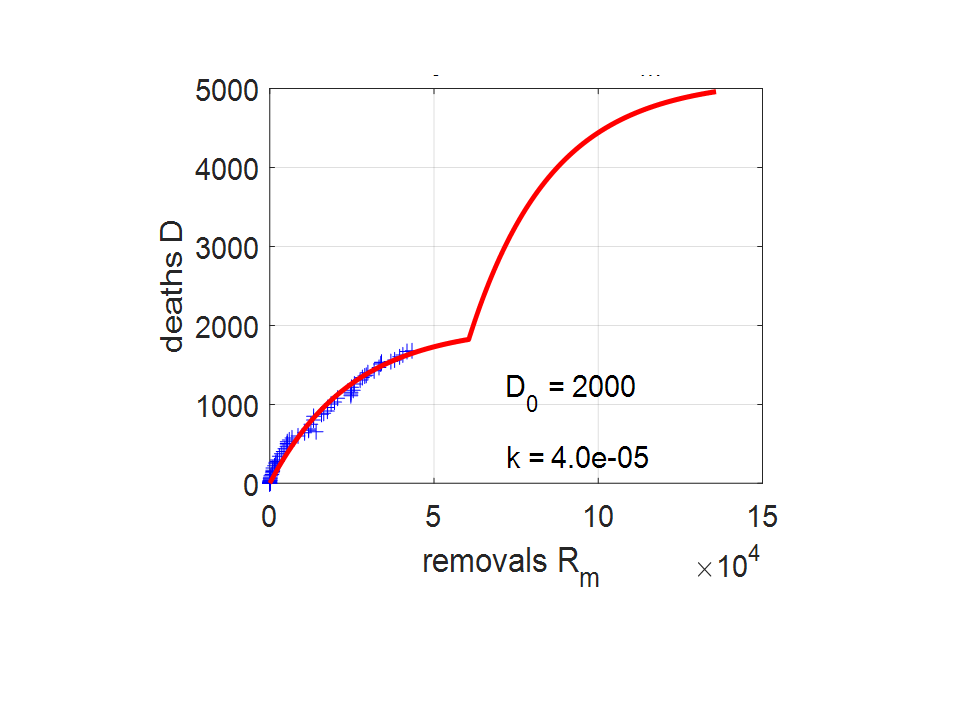}
	\caption{Texas: If a second wave occurs, there could be increase in the number of deaths, $D $.}\label{Fig7A}
\end{figure}

The plots in Figs. \ref{Fig6} and \ref{Fig6A} show that the peak in the total cumulative number of infections has not been reached as early as June, however, the peak is probably not far away. If there are no surges in the susceptible population, then one could expect that by late September, 2020, the number of infections will have fallen to very small numbers and the virus will have been well under control with the total number of deaths in the order of 2 000.
In mid-May, 2020, some restrictions have been lifted in the state of Texas. The SIR model can be used to model some of the possible scenarios if the early relaxation of restrictions leads to increasing number of susceptible populations. If there is a relatively small increase in the future number of susceptible individuals, no series impacts occur. However, if there is a large outbreak of the virus, then the impacts can be dramatic. For example, at the end of June, 2020, if $S$ was reset to 0.8 $(S = 0.8)$, a second wave of infections occurs with the peak number of infections occurring near the end of July, with the second wave peak being higher than the initial peak number of infections. Subsequently, the number of deaths will rise from about 2 000 to nearly 5 000, as shown in Figs. \ref{Fig7} and \ref{Fig7A}.

If governments start lifting their containment strategies too quickly, then it is probable there will be a second wave of infections with a larger peak in active cases, resulting to many more deaths.
\subsection{Italy}

\begin{figure}
	\centering
	\includegraphics[width=14cm,height=8.5cm]{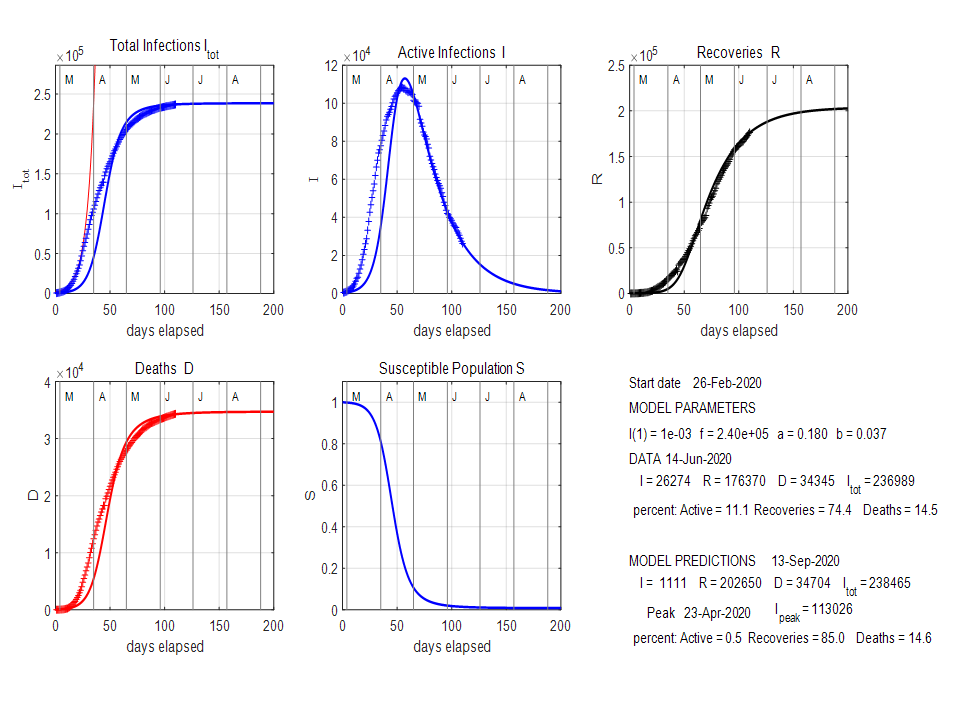}
	\caption{Italy: Model predictions for the period from 26 February to 13 September, 2020 with data from February to June, 2020.} 
	\label{Fig8}
\end{figure}
\begin{figure}
	\centering
	\includegraphics[width=12cm,height=8.5cm]{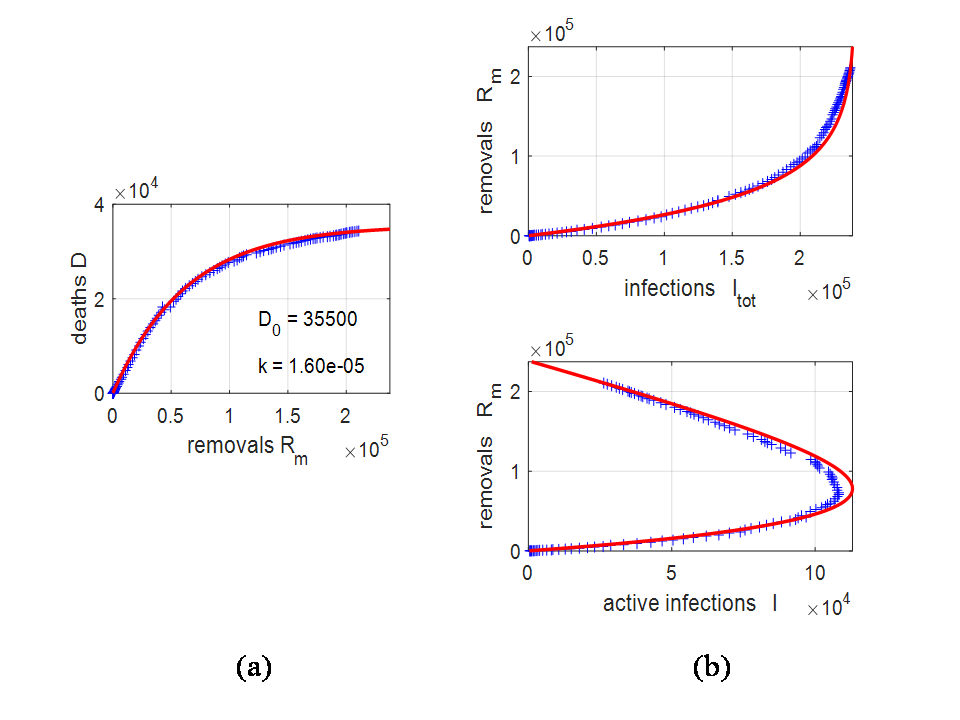}
	\caption{Italy: (a) Nonlinear fitting with Eq. \eqref{nonlinear_fitting_function_Dd_Rmd} using a trial-and-error method to estimate the number of deaths, $D$ from the removed population, $R_m$ (see text for the details). (b) Plots of the number of removals, $R_m$ against the cumulative total infections $I_{tot}$ and current active cases $I$.} 
	\label{Fig8A}
\end{figure}
Figure \ref{Fig8} shows clearly that the peak of the pandemic has been reached in Italy and without  further surge periods, the spread of the virus is contained and number of active cases is declining rapidly. The plots in panels (a), (b) in Fig. \ref{Fig8A} are a check on how well the model can predict the time evolution of the virus. These plots also assist in selecting the model's input parameters.

\section{Flattening the curve}\label{sec_flattening_the_curve}

\begin{figure}
	\centering
	\includegraphics[width=14cm,height=8.5cm]{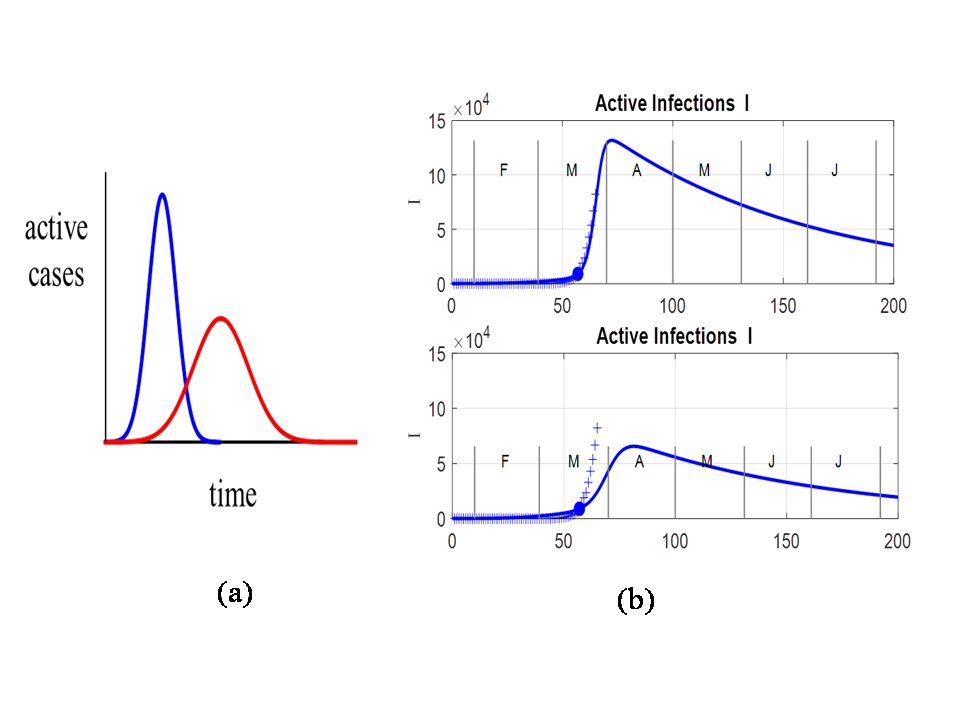}
	\caption{Flattening the curve: Panel (a): The {\it flattening of the curve} diagram used widely in the media to represent a means of reducing the impacts of COVID-19. Panel (b) If the number of susceptible individuals is reduced, then the peak number of infections will be less and the time for the number of infections to fall to low numbers is reduced.}
	\label{Fig9}
\end{figure}

The term {\it flattening the curve} has rapidly become a rallying cry in the fight against COVID-19, popularised by the media and government officials. Claims have been made that flattening the curve results in: (i) reduction in the peak number of cases, thereby helping to prevent the health system from being overwhelmed and (ii) in an increase in the duration of the pandemic with the total burden of cases remaining the same. This implies that social distancing measures and management of cases, with their devastating economic and social impacts, may need to continue for much longer. The picture which has been widely shown in the media is shown in Fig. \ref{Fig9}(a). 

The idea presented in the media as shown in Fig. \ref{Fig9}(a) is that by flattening the curve, the peak number of infections will decrease, however, the total number of infections will be the same and the duration of the pandemic will be longer. Hence, they concluded that by {\it flattening the curve}, it will have a lesser impact upon the demands in hospitals. Figure \ref{Fig9}(b) gives the scientific meaning of {\it flattening the curve}. By governments imposing appropriate measures, the number of susceptible individuals can be reduced and combined with isolating infected individuals, will reduce the peak number of infections. When this is done, it actually shortens the time the virus impacts the society.
Thus, the second claim has no scientific basis and is incorrect. What is important is reducing the peak in the number of infections and when this is done, it shortens the duration in which drastic measures need to be taken and not lengthen the period as stated in the media and by government officials. Figure \ref{Fig9} shows that the peak number of infections actually reduces the duration of the impact of the virus on a community.

\section{Conclusions}\label{sec_conclusions}
Mathematical modelling theories are effective tools to deal with the time evolution and patterns of disease outbreaks. They provide us with useful predictions in the context of the impact of intervention in decreasing the number of infected-susceptible incidence rates \cite{Giordano2020,Hou2020,Anas2020}.

In this work, we have augmented the classic SIR model with the ability to accommodate surges in the number of susceptible individuals, supplemented by recorded data from China, South Korea, India, Australia, USA and the state of Texas to provide insights into the spread of COVID-19 in communities. In all cases, the model predictions could be fitted to the published data reasonably well, with some fits better than others. For China, the actual number of infections fell more rapidly than the model prediction, which is an indication of the success of the measures implemented by the Chinese government. There was a jump in the number of deaths reported in mid-April in China, which results in a less robust estimate of the number of deaths predicted by the SIR model. The susceptible population dropped to zero very quickly in South Korea showing that the government was quick to act in controlling the spread of the virus. As of the beginning of June, 2020, the peak number of infections in India has not yet been reached. Therefore, the model predictions give only minimum estimates of the duration of the pandemic in the country, the total cumulative number of infections and deaths. The case study of the virus in Australia shows the importance of including a surge where the number of susceptible individuals can be increased. This surge can be linked to the arrival of infected individuals from overseas and infected people from the Ruby Princess cruise ship. The data from USA is an interesting example, since there are multiple epicentres of the virus that arise at different times. This makes it more difficult to select appropriate model parameters and surges where the susceptible population is adjusted. The results for Texas show that the model can be applied to communities other than countries. Italy provides an example where there is excellent agreement between the published data and model predictions.

Thus, our SIR model provides a theoretical framework to investigate the spread of the COVID-19 virus within communities. The model can give insights into the time evolution of the spread of the virus that the data alone does not. In this context, it can be applied to communities, given reliable data are available. Its power also lies to the fact that, as new data are added to the model, it is easy to adjust its parameters and provide with best-fit curves between the data and the predictions from the model. It is in this context then, it can provide with estimates of the number of likely deaths in the future and time scales for decline in the number of infections in communities. Our results show that the SIR model is more suitable to predict the epidemic trend due to the spread of the disease as it can accommodate surges and be adjusted to the recorded data. By comparing the published data with predictions, it is possible to predict the success of government interventions. The considered data are taken in between January and June, 2020 that contains the datasets before and during the implementation of strict and control measures. Our analysis also confirms the success and failures in some countries in the control measures taken.

Strict, adequate measures have to be implemented to further prevent and control the spread of COVID-19. Countries around the world have taken steps to decrease the number of infected citizens, such as lock-down measures, awareness programs promoted via media, hand sanitization campaigns, etc. to slow down the transmission of the disease. Additional measures, including early detection approaches and isolation of susceptible individuals to avoid mixing them with no-symptoms and self-quarantine individuals, traffic restrictions, and medical treatment have shown they can help to prevent the increase in the number of infected individuals. Strong lockdown policies can be implemented, in different areas, if possible. In line with this, necessary public health policies have to be implemented in countries with high rates of COVID-19 cases as early as possible to control its spread. 	
The SIR model used here is only a simple one and thus, the predictions that come out might not be accurate enough, something that also depends on the published data and their trustworthiness. However, as the model data show, one thing that is certain is that COVID-19 is not going to go way quickly or easily.

\section*{Acknowledgements}
AM is thankful for the support provided by the Department of Mathematical Sciences, University of Essex, UK to complete this work.


\end{document}